\documentclass[prd,aps,12pt,nofootinbib,reprint]{revtex4-1} % article
\usepackage{mathrsfs,amssymb,amsmath,amsxtra,amsfonts}
\usepackage{graphicx,color}
\usepackage{verbatim}
\usepackage[colorlinks]{hyperref}
\usepackage[utf8x]{inputenc}

\begin{document}

\newcommand{\g}{\mbox{$\mathfrak{g}$}}
\newcommand{\x}{\mbox{$\mathfrak{x}$}}
\renewcommand{\r}{\mbox{$\mathfrak{r}$}}
\renewcommand{\t}{\mbox{$\mathfrak{t}$}}

%opening
\title{Do ghost fields really make massive gravity invalid?}
\author{I. Schmelzer}
\noaffiliation

\begin{abstract}
Theories which contain ghost fields are considered to be invalid. It is assumed that for such theories the energy is unbounded from below, and the theory will be unstable, allowing the creation of particle pairs with arbitrarily large positive and negative energies. Such ghost fields are a serious problem for theories of massive gravity.

We present here an example of such a theory of gravity with a ghost field which avoids this problem. While its equations are locally equivalent to those of a theory of massive gravity with a ghost field, global restrictions of the configuration space have the result that the ghost mode has to remain small, and the energy has to remain positive.

As a consequence, the existence of a ghost-like mode in the equations of the theory is not a decisive argument against this theory.
\end{abstract}
\maketitle

\section{Introduction}

Recent observations of gravitational waves from black hole mergers have found ``tentative evidence for Planck-scale structure near black hole horizons'' \cite{abyss,abyss3}. This would make theories which predict stable gravastars slightly greater than horizon size instead of GR black holes more interesting. Such stable gravastars are known to appear in theories of massive gravity.

Unfortunately, many theories of massive gravity have serious problems with ghost fields -- fields with negative energy, which would make the theory unstable. For a relativistic massive spin 2 particle, in the linear approximation the most general mass term is \(\mathcal{L}_{{\mathrm  {int}}}=ah^{{\mu \nu }}h_{{\mu \nu }}+b\left(\eta ^{{\mu \nu }}h_{{\mu \nu }}\right)^{2}\), where  \( g_{\mu \nu }=\eta _{\mu \nu }+M_{\mathrm {Pl} }^{-1}h_{\mu \nu }\). Fierz and Pauli \cite{FierzPauli} showed in 1939 that the only possibility to avoid the ghost mode is \(a=-b\). This theory has another problem, the vDVZ discontinuity \cite{vDV,Zakharov}. Moreover, Boulware and Deser \cite{Boulware} have argued that the ghost will reappear if the non-linear terms are considered. Recently, possibilities to circumvent these problems have been proposed \cite{deRham,deRham2,Hassan}. Nonetheless, these are exceptions rather than the rule, and ghost fields are present in many theories of massive gravity.

But is a ghost field really as fatal as it seems at a first look? The aim of this paper is to present an example of a theory -- General Lorentz Ether (GLE) \cite{clm,glet} -- which is for some parameter choices equivalent to a theory of massive gravity with a ghost field. We will show here that the ghost field in this theory appears to be completely harmless.

The main result is that the configuration space of the theory simply does not contain field configurations with large ghost field contributions, so that the energy in the theory remains bounded from below everywhere. This makes the ghost field unproblematic in principle. The cause of this restriction of the configuration space is the very construction of the theory: It is defined as a condensed matter theory on a Newtonian background, and the interpretation as a theory of massive gravity -- a metric theory of gravity with four additional degrees of freedom -- is based on a non-trivial embedding of the space of valid configurations of the theory into the larger space of a metric theory of gravity with four additional scalar fields. The problematic configurations of the metric theory with large ghost contributions lie outside the image of this embedding, thus, do not define field configurations of the theory itself. In particular, the ghost mode is simply the Newtonian time coordinate of the theory, and has to be time-like. This property would be destroyed by large ghost mode contributions.

We extend the result to another theory of massive gravity, the ``Relativistic Theory of Gravity'' (RTG). For some parameter settings, the equations are equivalent to GLE. We find that the ``causality condition'' used in RTG has a similar restricting effect, so that the ghost field is unproblematic in RTG with causality condition too. This strengthens some results \cite{Loskutov,GershteinFlow} that, despite the ghost mode, the gravitational radiation of stars is positive definite.

We consider also the case of other theories of massive gravity. The observation that most of the considerations do not depend on the particular equations of the theory, but only on the interpretation, namely that the ghost field $\t(x)$ is interpreted as a preferred time coordinate and therefore has to be time-like, makes it plausible that ghost field problems may be solved for other theories of massive gravity in a similar way.

\section{The General Lorentz Ether}

The theory of gravity which allows to circumvent the ghost problem has been proposed in \cite{clm}, \cite{glet}. It is a theory of gravity with a condensed matter interpretation of the gravitational field, which generalizes the Lorentz ether interpretation of special relativity and has, therefore, be named ``General Lorentz Ether'' (GLE). The theory has a Newtonian background of absolute space and time, the preferred coordinates we will denote by \(\x^\mu\). The Lagrangian of the theory is
\begin{equation}\label{S}
L = -\frac{1}{16\pi G}
       	(\Upsilon g^{00}-\Xi g^{ii})\sqrt{-g}
         + L_{GR} +  L_{matter}
\end{equation}
It appears useful to introduce a formally covariant form of the Lagrangian where the preferred coordinates are represented by four scalar functions \(\x^i(x), \x^0(x)=\t(x)\) of general coordinates \(x\).  This gives
\begin{equation}\label{Scov}
\begin{split}
L = &-\frac{1}{16\pi G}
       	(\Upsilon g^{\mu\nu}\frac{\partial\t}{\partial x^\mu}\frac{\partial\t}{\partial x^\nu}-\Xi g^{\mu\nu}\frac{\partial\x^i}{\partial x^\mu}\frac{\partial\x^i}{\partial x^\nu})\sqrt{-g}\\
         &+ L_{GR} +  L_{matter}
\end{split}
\end{equation}
So, the preferred coordinates look here like scalar fields in a formally covariant Lagrangian. The Euler-Lagrange equations therefore give the harmonic wave equation for the ``fields'' \(\x^\alpha(x)\)
\begin{equation}
\square \x^\alpha(x) = 0
\end{equation}
which in the preferred coordinates gives
\begin{equation} \label{harm}
 \frac{\partial}{\partial \x^\mu} (g^{\mu\alpha}(\x,\t)\sqrt{-g}) = 0.
\end{equation}
In a variant of the ADM decomposition \cite{ADM}, the preferred coordinates $\x^i$, $\t$ split the gravitational field $g^{\mu\nu}\sqrt{-g}$ into a scalar $\rho$, a three-vector $v^i$ and a three-metric $\sigma^{ij}$:
\begin{subequations}\label{gdef}
\begin{align}
g^{00}\sqrt{-g} &= \rho, \\
g^{0i}\sqrt{-g} &= \rho v^i, \\
g^{ij}\sqrt{-g} &= \rho v^i v^j - \sigma^{ij}.
\end{align}
\end{subequations}
so that the harmonic condition \eqref{harm} splits into the following equations:
\begin{subequations}\label{classical}
\begin{align}
\label{continuity}
\partial_t \rho + \partial_i (\rho v^i) &= 0, \\
\label{Euler}
\partial_t (\rho v^j) + \partial_i(\rho v^i v^j - \sigma^{ij}) &= 0.
\end{align}
\end{subequations}
which have a natural interpretation as continuity and Euler equations of a classical condensed matter theory.

\section{GLE as a theory of massive gravity with ghost mode}

The equations of the theory itself appear, for the particular choice of the signs of the free parameters as $\Xi,\Upsilon>0$ and $\Lambda<0$, to be equivalent to those of the ``relativistic theory of gravity'' (RTG).\footnote{This theory has been first proposed by \mbox{Freund}, Maheshwari and Schonberg 1969 \cite{FMS} and appears in a two-parameter family of Lagrangians constructed by Ogievetsky and Polubarinov 1965 \cite{OP} as the case $q=0$, $p=-1$.  Later, the Lagrangian has been rediscovered by Logunov and coworkers \cite{Logunov,Logunov1,Logunov2,Logunov3}, who have named the theory ``relativistic theory of gravity'' (RTG). Because this name is widely used and neutral we will use it too (instead of, say, FMS theory). For an introduction into RTG see \cite{Logunov2}.} RTG is a bimetric theory of massive gravity, with a Minkowski background metric \(\eta^{\mu\nu}\), where the mass m and the coefficients of the background metric are connected with the coefficients in \eqref{S} by
\begin{equation}
        \Lambda=-\frac{m^2}{ 2}<0,
        \Xi=-\eta^{11}\frac{m^2}{ 2}>0,
        \Upsilon=\eta^{00}\frac{m^2}{ 2}>0.
\end{equation}
The choice \(\Upsilon > 0\) not only makes the theory equivalent to RTG, it is also the obviously ``wrong'' sign if we look at the four preferred coordinates as if they were massless scalar fields. This wrong sign makes the preferred time coordinate \(\t(x)\) a ghost mode. But it is also this ``wrong'' sign which causes the stop of the gravitational collapse and gives stable gravastars with a radius slightly greater than the Schwarzschild radius.

\section{The ghost mode is massless dark matter}

The first observation is that the ghost mode is ideal dark matter. In the theory, the EEP holds exactly, so that the matter Lagrangian cannot depend on the preferred coordinates. But the ghost mode is the preferred time coordinate. So, any interaction between usual matter and the ghost mode is forbidden by the EEP. The only way the ghost mode may interact with anything is, therefore, the gravitational interaction.

Moreover, the ghost mode is massless, thus, if ghost particles would be created at some point, they would be radiated away immediately, thus, would show up in the gravitational radiation. Here we could refer to the results of \cite{Loskutov,GershteinFlow} that, despite the ghost mode, the gravitational radiation of stars is positive definite.

The only place where they could stay bounded despite being massless would be near the Schwarzschild radius. An effect of the additional term near the Schwarzschild radius exists, indeed, and is well-studied -- namely, it stops the gravitational collapse and leads to stable gravastars, which is what makes the theory interesting in the light of \cite{abyss,abyss3}.  So, already a rough overview suggests that the ghost mode is the most harmless ghost mode imaginable.

\section{Positive energy for classical solution}

Let's now take a look at the energy-momentum conservation laws. There are two variants of the energy conservation law. The first one can be formally obtained as the Euler-Lagrange equation for the preferred coordinates, $\frac{\delta S}{\delta \t} = \square \t = 0$, which give the harmonic condition \eqref{harm} and the full energy-momentum tensor

\begin{equation}\label{emt}
T^{\mu\nu}\sqrt{-g} = g^{\mu\nu}\sqrt{-g}
\end{equation}
In this form, it would not depend on matter fields at all. But the traditional form of the energy-momentum tensor $T^{\mu\nu}_{gravity} + T^{\mu\nu}_{matter}$ can be easily obtained too, using the Euler-Lagrange equation for the \(g_{\mu\nu}\). The point is that, despite the additional term, the equations of the theory have a form similar to the Einstein equations, $T_{matter}^{\mu\nu} = G^{\mu\nu} + F(g^{\mu\nu},\x^\alpha)$, where the additional term depends only on the gravitational field and the preferred coordinates, but not on matter fields. This allows us to add zero to the energy-momentum tensor
\begin{equation}
\begin{split}\label{emtsplit}
T^{\mu\nu}\sqrt{-g} &= g^{\mu\nu}\sqrt{-g} + T^{\mu\nu}_{matter} - G^{\mu\nu} - F(g^{\mu\nu},\x^\alpha)\\
&=  T^{\mu\nu}_{matter} +  T^{\mu\nu}_{grav}
\end{split}
\end{equation}
So, we have classical energy-momentum conservation, and, given the interpretation of $T^{00}\sqrt{-g}=\rho\ge 0$ as the density of the Lorentz ether, the energy density is non-negative everywhere.

\section{What makes the ghost mode compatible with positive energy}

To get a better understanding how the presence of a ghost mode may be compatible with an energy density which is always non-negative, one has to look at the way how the ghost mode has appeared in the theory.

In the formulation of the theory in the preferred coordinates, the time coordinate is not a variable at all. What defines the energy density is the density of the Lorentz ether, and it is always positive. Configurations with negative ether density simply do not exist.

Then we construct an embedding of all valid configurations of the Lorentz ether into the much larger space of configurations of a metric theory of gravity with four scalar fields $\x^\alpha(x)$, by using the preferred coordinates as these scalar fields. This is an embedding: Every valid field configuration with a density $\rho(\x,\t)>0$ defines a metric $g^{\mu\nu}(x)$ as well as four scalar fields $\x^\alpha(x)$, but not every such field configuration is the result of such an embedding.

This embedding of the original condensed matter theory into a metric field theory with scalar fields is quite useful as a mathematical tool, and almost necessary for comparison of the predictions of the theory in comparison with GR. But the theory is defined as a condensed matter theory, and not as metric theory of gravity with scalar fields.

As a condensed matter theory, the density will be always non-negative, automatically.  So, only solutions of the field theory with $g^{00}(x)\sqrt{-g}\ge 0$ can be images of valid Lorentz ether configurations.  So, while (given the ghost mode) the metric theory of gravity with scalar fields will have an energy which is locally not bounded from below, the subspace of  images of valid Lorentz ether configurations will have an energy bounded from below everywhere.

\section{What happens if the metric theory is unstable}

One can nonetheless argue that a ghost instability would be problematic anyway, simply because the theory would become invalid. Say, we start with a valid initial configuration, but the unstable evolution leads after some time to a state which is no longer a valid field configuration.

In this case, the solution becomes invalid at the moment where the density of the Lorentz ether reaches zero. Up to this moment, the solution is a meaningful solution. The moment itself has also a meaningful interpretation:  The ether tears into parts, and there appears a region in the background space where the ether density is zero. It is obvious that the region where the ether density is zero is no longer correctly described by the equations. So, we have new physics. In fact, we need new physics at the very moment where the ether density becomes zero -- starting with this moment, we would need a variant of the condensed matter theory where the ether has a border, with some well-defined boundary conditions.

The very possibility that condensed matter may tear into parts is well-know for usual condensed matter, and, in itself, does not mean that the condensed matter theory itself is somehow wrong. Simply if this happens, the domain of applicability of the condensed matter theory, which is applicable only for $\rho>0$, has been reached.

It is worth to note that this is the way how the possibility of solutions with closed causal loops is handled in GLE. For a solution with a closed causal loop, no global time-like coordinate exist, thus, for every time coordinate where would be a region where $g^{00}\sqrt{-g} = \rho < 0$. So, valid GLE configurations cannot contain causal loops, and the preferred time coordinate is always time-like.

\section{What about quantum fluctuations?}

One could argue that, even if in classical theory everything seems fine, one cannot be sure that something bad happens in quantum theory. Let's see.

First of all, the theory one would have to quantize would be the condensed matter theory, and not the metric theory with scalar fields. Indeed, quantum fluctuations of the ``fields'' \(\x^\mu(x)\) would not obliged to define valid global coordinates, thus, would not be part of the image of condensed matter theory. Thus, the embedding into the field theory is simply not an appropriate tool to handle the quantum theory.

So, we would have to consider quantization in the space of valid configurations of the Lorentz ether, and not for the metric theory with scalar fields. And, whatever the quantization procedure in this space, the resulting expectation values have to be expectation values of meaningful Lorentz ether configurations, thus, there has to be $\langle \psi | \hat{\rho}(\x) | \psi\rangle \ge 0$. Quantum fluctuations can violate classical evolution equations, but nonetheless remain reasonable configurations, from the same classical configuration space $Q$.

The straightforward way to quantize a condensed matter theory gives even more rigorous restrictions. What one would quantize would be an atomic ether variant, which could be modeled by a co-moving lattice, so that the density $\rho(\x)$ describes the density of lattice nodes and the nodes would move with the velocity $v^i(\x)=g^{0i}/g^{00}$. In this case, the continuity equation \eqref{continuity} would become a tautology. Then, it would be the position of the lattice nodes (and not their number) which would be quantized. As a consequence, the continuity equation would not obtain any quantum fluctuations, as well as the integral $\int \rho(\x) d^3\x$ over the whole space would remain constant, without any quantum fluctuations.

Instead, to quantize a condensed matter theory as a field theory by postulating some commutation relations as canonical can lead to problems, like the possibility of negative densities. But this would be a problem of a wrong approach to quantize the condensed matter theory, not a problem of the condensed matter theory itself.

So, even quantization will not endanger the basic property $T^{00}\sqrt{-g}=\rho\ge 0$, so that the ghost mode remains harmless even in the quantum domain. Let's not that this consideration remains valid also for an arbitrary matter Lagrangian. Even if we would allow violations of the EEP, either because of quantum theory making EEP violations necessary, or because of the necessity of modifications of the theory at the boundary of the ether, this would not change the methods described here, and the consequences.  If the quantum ether tears into parts, the nodes concentrate in one part, leaving large parts without any nodes, but the number of nodes remains fixed.

\section{But maybe we need a different notion of energy?}

We have obtained the energy-momentum conservation law \eqref{emtsplit} in its usual form by adding the equations of motion to \eqref{emt}. What we have considered up to now is the positivity of the Lorentz ether density $g^{00}\sqrt{-g}$, so, of the energy as defined by \eqref{emt}.  But maybe we need in quantum theory the energy in the usual form, thus, as
\eqref{emtsplit}? In this case, the fact that the derivation has used the classical equation of motion makes it invalid in the quantum case, and $H=T^{00}>0$ would hold only in the classical limit.  So, maybe a high momentum quantum fluctuation of the ghost mode nonetheless allows to reach large negative energies?

Let's start with an arbitrary valid configuration of the theory. As the matter configuration, as the metric may be arbitrary (so we can assume starting from some quantum fluctuation, and allow for quantum EEP violations), except that they have to define a valid field configuration, thus, a positive  Lorentz ether density \(\rho(\x) > 0\), or $\t(x)$ being a time-like coordinate. Let's now consider some local disturbance of the ghost mode \(\delta \t(x)\). To define a valid coordinate \(\t'(x)=\t(x)+\delta \t(x)\), the disturbance  \(\delta \t(x)\) has to be smooth. Assume it is nonzero at some event \((\x_e,\t_e)\). Given that it is a local disturbance, it will increase from as well as decrease to zero somewhere on the line \((\x_e,\t)\). So there will be moments \(\t_+\) where it increases, \(\partial_{\t} \delta\t(\x_e,\t_+) > 0\), resp. \(\t_-\) where it decreases in time \(\partial_{\t} \delta\t(\x_e,\t_-) < 0\).

Let's now consider how the coordinate behaves in dependence of a parameter c, with \(\t_c = \t + c\delta\t\). For \(c\) sufficiently small, \(\t_c\) will remain a valid time-like coordinate. But for c becoming large enough, the time derivative at \(\t_-\) will become negative: \(\partial_{\t} \delta\t_c(\x_e,\t_-) < 0\). On the other hand, at \(\t_+\) the time derivative remains positive: \(\partial_{\t} \delta\t_c(\x_e,\t_+) > 0\). As a consequence, there will be some intermediate moment of time \(\t_0\) between them where \(\partial_{\t} \delta\t_c(\x_e,\t_0) = 0\). If for all the other \(\partial_i \t_c(\x_e,\t_0)=0\) too, \(\nabla \t_c(x)=0\) and the Jacobian will be zero, so that \(\t_c(x)\) cannot be used to define a valid system of coordinates. Else, at \((\x_e,\t_0)\) the gradient \(\nabla \t_c\) gives zero for some time-like vector, namely the tangential vector of \((\x_e,\t)\). But this is what characterizes a space-like function. Thus, the time coordinate \(\t_c\) defines a configuration with negative density \(\rho(\x_e,\t_0)<0\), and, therefore, not a valid configuration of the theory.

Note that this consideration excludes, together with some particular value of \(c=c_0\), as well all greater values \(c>c_0\), because \(\partial_{\t} \delta\t_c(\x_0,\t_-) < 0\) remains negative if we increase \(c\). Note also that the analogical consideration will exclude also negative values of \(c\) with \(|c|\) large enough to make \(\partial_{\t} \delta\t_c(\x_e,\t_+) < 0\).  So, for each ghost mode there exists only a finite range \(|c|< c_0\) where the configuration can define a valid configuration of the theory.

So we conclude that the possible contributions of the ghost mode itself are restricted, independent of any equations of motion or particular assumptions about the energy.

\section{The maximal contribution of the ghost mode to energy}

What makes us think that the ghost mode is problematic is, of course, that we have an assumption about the energy of the ghost mode, namely that of the linear approximation of the field theory. This makes us believe that especially the high momentum contributions -- those with large negative energy -- would be problematic. To consider the question if these hing momentum contributions are possibly problematic, we can restrict ourselves to the simplest case of a Minkowski background. In this case, the energy of a scalar ghost field would be
\begin{equation}
H= - \int d^{3}\x\left[\frac12 (\partial_{\t} \varphi)^{2}+\frac12(\nabla \varphi)^{2}\right].
\end{equation}
Our restriction for \(\delta \t\) is that \(|\partial_{\t} \delta\t| < 1\). If we restrict ourself to pure modes of fixed momentum, this restriction would also lead to the additional restrictions  \(\partial_{\t} \delta\t < 1\) and \(|\nabla \delta\t| < 1\). Combining different modes, one can, of course, easily construct solutions which locally violate these restrictions. Nonetheless, a restriction valid for all modes with definite momentum can be considered as a reasonable bound for the general case. So, with these additional restrictions we obtain
\begin{equation}
H > - \int d^3 \x.
\end{equation}
This is a quite simple infrared infinity. Once such an infrared regularization is anyway necessary to obtain finite values for energy, this is unproblematic. The standard regularization by  introducing a finite volume, say, a large cube with periodic boundary conditions, regularized the maximal contribution of the ghost mode too. It appears to be proportional to the volume \(V\) of the cube.

Note that this bound for the energy does not depend on the momentum of the ghost mode. Once the bound, restricted to a finite volume, is finite, it follows that for a too large momentum even a single ghost particle would have an energy outside this bound. Thus, there is also a corresponding bound for the maximal possible momentum of a ghost mode. So, ghost modes with larger momentum cannot even be excited, because the average field configuration of even the first excited state would be an invalid configuration.

As a consequence, we can also forget about the danger that even a single quantum jump could create a ghost with very high momentum together with some usual matter and cause some problems. Ghost particles with too high momentum simply do not exist in the subspace of valid configurations.

\section{Does this result extend to RTG?}

In RTG, there is no condensed matter interpretation, and therefore also no condition that the ether density has to be positive.

But in RTG there is also a quite similar additional condition, the RTG causality condition. It requires that the light cone of the effective, observable metric \(g_{\mu\nu}(x)\) is inside the light cone of the background metric \(\eta_{\mu\nu}\). This is justified by the hypothesis that the background metric is physically fundamental, so that Einstein causality holds in relation to the background metric. It follows that the observable Einstein causality restricted by the observable metric cannot violate the fundamental Einstein causality of the background metric.

The causality condition is even more restrictive than the condition that the GLE density should be positive. Indeed, if the light cone of the effective physical metric is inside the light-cone of the background Minkowski metric, the time coordinate of the Minkowski metric will be also a time-like coordinate of the physical metric. That means, if we accept the causality condition as an acceptable part of RTG, then the considerations above show also that in RTG the energy will be bounded below too.

On the other hand, the equations of RTG clearly have solutions which violate the causality condition. How this conflict between the equations of the theory and the causality condition is handled remains unclear. In comparison, the situation in GLE is clear -- solutions where the ether density becomes negative make no physical sense, the equations become invalid if the ether density becomes zero and one has to modify the theory by introducing boundary conditions where the ether density becomes zero. Instead, what happens in this case in RTG remains undefined.

So, while the causality condition solves the problem with the ghost field, there is a conflict between the evolution equations of the theory and the causality condition. Valid initial conditions can evolve into configurations which violate the causality condition.

If one solves this conflict by rejecting the causality condition, the ghost problem reappears. In this case the properties of the ghost field -- that it is ideal massless dark matter -- can nonetheless be used to argue that the ghost mode is harmless in RTG too.

\section{Can the result be expected to be even more general?}

To understand if analogical considerations may be applicable also to other theories of massive gravity with ghost fields, let's have a look at the basis. A theory of massive gravity would not be background-independent like GR. So, there would be four additional degrees of freedom related with the choice of preferred coordinates \(\x^\mu(x)\). In theories of massive gravity, the additional degrees of freedom, which include the ghost, are those four additional degrees of freedom related with the background. Moreover, once can also expect that the ghost mode is closely connected with the time coordinate too.

The main argument depends only on the time coordinate becoming space-like in the background metric for large values of the ghost mode. If this makes the solution invalid or not depends on the physical interpretation of the theory. The condition of non-negativity of an ether density as well as the causality condition of RTG where the light cone of the effective metric has be remain inside the light cone of the background metric are examples of interpretations which exclude such solutions from the theory. Other interpretations may not care about this.  In this case, the ghost mode defines valid solutions of the theory even if large, the argument is not applicable, and the ghost problem remains.

Nonetheless, given that up to now the consideration of interpretations of field theories was widely ignored, based on the prejudice that they have no physical consequences, one cannot exclude in any way that there exist reasonable interpretations of other theories of massive gravity which exclude large ghost modes in a similar way. Moreover, most of the considerations given here do not depend on the particular choice of the equations, but rely completely on the assumption that the ghost mode $\t(x)$ has to be interpreted as a preferred time-like coordinate.

So, the considerations given here open a quite general way to handle the Boulware-Deser ghost field in theories of massive gravity -- to interpret the ghost mode as a preferred coordinate, which has to be time-like.

\section{Conclusion}

We have found a surprising result, namely that the ghost fields, which seem to invalidate most theories of massive gravity, may appear completely unproblematic. In particular, we have studied the example of GLE, a generalization of the Lorentz ether to gravity, which, for a particular choice of the free parameters of the theory $\Upsilon>0$, obtains a ghost mode. It appears that the theory has nonetheless not only a global positive energy, but even an energy density positive everywhere.

While the ghost mode of the theory is, from the start, of the most harmless type imaginable (interaction with matter is strictly forbidden by the EEP, and the mode itself is massless), the energy density would remain non-negative even if we would allow EEP-violating interactions with matter and quantum fluctuations. The result is rigorous, because it is based on the consideration of the configuration space of the theory itself. The field-theoretic description as a metric theory with some additional scalar fields is only the result of a non-trivial embedding of the configuration space of the theory into the much larger configuration space of the metric theory. All solutions with large contributions of the ghost mode appear to be outside the image of the configuration space of the original theory, thus, do not define configurations of the original theory.

In particular, we have considered the maximal possible contribution of a general ghost mode, and found that after an infrared regularization, which restricts the mode to a finite volume, we obtain a maximal possible contribution of the ghost mode which does not depend on its momentum.  As a consequence, ghost particles with too large momentum simply do not exist in the quantum theory.

We have also considered the question if similar considerations can be extended to other theories of massive gravity. We have found that, in particular, in RTG the causality condition has a similar effect, so that the ghost mode is unproblematic in RTG too.  In general, we have concluded that the key result did not depend on the particular Lagrangian of the theory. The key was that the fundamental space of field configurations of the theory was not the full space of all metrics, together with all possible configurations of the ghost field, but restricted by the interpretation.

We have found it plausible that a similar interpretation of the ghost field as defining a preferred time-like time coordinate $\t(x)$ could have similar effects for other theories of massive gravity too. So, a reinterpretation of the theory can solve the problem of ghost fields for other theories of massive gravity too.

The result is also interesting from a general, methodological point of view: The same set of field equations may appear, in one interpretation, to be a viable theory with positive energy, but in another interpretation invalid because of an unrestricted ghost mode. So, interpretations of physical theories matter, and sometimes in very unexpected ways. Once an interpretation can solve such a serious problem like that of ghosts, maybe they are able to solve other problems, for example naturalness problems, too?

\end{document}